\documentclass[aps,prb,twocolumn,superscriptaddress]{revtex4-2}

\usepackage{graphicx}
\usepackage{tabularx}
\usepackage{color}

\begin{document}


\title{Laser-driven shock compression and equation of state of Fe$_2$O$_3$ up to 700 GPa}



\author{Alexis Amouretti}
\affiliation{Institut de Min\'eralogie de Physique des Mat\'eriaux et de Cosmochimie, CNRS, Sorbonne Universit\'e, MNHN, Paris, FRANCE}
\affiliation{Graduate School of Engineering, Osaka University, Suita, Osaka, JAPAN}
\author{Marion Harmand}
\affiliation{Institut de Min\'eralogie de Physique des Mat\'eriaux et de Cosmochimie, CNRS, Sorbonne Universit\'e, MNHN, Paris, FRANCE}
\affiliation{PIMM, Arts et Metiers Institute of Technology, CNRS, Cnam, HESAM University, 75013 Paris, FRANCE}
\author{Bruno Albertazzi}
\affiliation{LULI, \'Ecole Polytechnique, CNRS, CEA, UPMC, Palaiseau, FRANCE}
\author{Antoine Boury}
\affiliation{Institut de Min\'eralogie de Physique des Mat\'eriaux et de Cosmochimie, CNRS, Sorbonne Universit\'e, MNHN, Paris, FRANCE}
\author{Alessandra Benuzzi-Mounaix}
\affiliation{LULI, \'Ecole Polytechnique, CNRS, CEA, UPMC, Palaiseau, FRANCE}
\author{D. Alex Chin}
\affiliation{Laboratory for Laser Energetics, University of Rochester, Rochester, UNITED STATES}
\author{Fran\c{c}ois Guyot}
\affiliation{Institut de Min\'eralogie de Physique des Mat\'eriaux et de Cosmochimie, CNRS, Sorbonne Universit\'e, MNHN, Paris, FRANCE}
\author{Michel Koenig}
\affiliation{LULI, \'Ecole Polytechnique, CNRS, CEA, UPMC, Palaiseau, FRANCE}
\author{Tommaso Vinci}
\affiliation{LULI, \'Ecole Polytechnique, CNRS, CEA, UPMC, Palaiseau, FRANCE}
\author{Guillaume Fiquet}
\affiliation{Institut de Min\'eralogie de Physique des Mat\'eriaux et de Cosmochimie, CNRS, Sorbonne Universit\'e, MNHN, Paris, FRANCE}


\date{\today}

\begin{abstract}
    We report here the first equation of state measurements of Fe$_2$O$_3$ obtained with laser-driven shock compression. The data are in excellent agreement with previous dynamic and static compression measurements at low pressure, and extend the known Hugoniot up to 700~GPa. We observe a large volume drop of $\sim$10\% at 86~GPa, which could be associated, according to static compression observations, with the iron spin transition. Our measurements also suggest a change of the Hugoniot curve between 150 and 250~GPa. Above 250~GPa and within our error bars, we do not observe significant modifications up to the maximum pressure of 700~GPa reached in our experiment. 
\end{abstract}


\maketitle

\section{Introduction}
Super-Earths are rocky exo-planets, with a mass 1 to 10 times that of the Earth. The increasing number of discoveries of such planetory bodies implies to extend the studies of the rocks composing them to much more important pressures than for the Earth \cite{valencia_internal_2006}. As a first approximation, the mantle of a Super-Earth is supposed to be similar to the Earth mantle, and essentially composed of Mg-rich silicates and oxides \cite{duffy_mineralogy_2015}. The presence of iron oxides, even in small quantities, plays a fundamental role in the formation of minerals in the Earth's mantle.
It is thus necessary to define the fundamental physical properties, starting with the equation of state (EOS) and phase transition boundaries, of iron oxides at very high pressure.  Among iron oxides, w\"ustite FeO and hematite Fe$_2$O$_3$ are the two end-members of the ferrous and ferric states of iron. Unlike FeO, studied in static and dynamic compression at very high pressure \cite{ozawa_phase_2011, coppari_implications_2021}, data on Fe$_2$O$_3$ are still limited to a rather small pressure range.

Several static compression studies agreed that Fe$_2$O$_3$ goes through a series of phase transitions at 50~GPa, including a Mott transition and a spin transition \cite{greenberg_pressure-induced_2018, pasternak_breakdown_1999, badro_nature_2002, sanson_local_2016}. In particular, the pressure-induced spin transition of iron is well known and it can significantly affect the physical and chemical properties of the mantle minerals \cite{badro_spin_2014}. 
Most recent studies have measured the isothermal compression of Fe$_2$O$_3$ up to 105~GPa \cite{bykova_structural_2016} using diamond anvil cells (DAC) and has been validated with a theoretical approach using density functional theory and dynamical mean-field theory (DFT + DMFT) up to 100~GPa \cite{greenberg_pressure-induced_2018}.
First, the rhomboedric $\alpha$-Fe$_2$O$_3$ phase is stable up to $\sim$~40~GPa. Between 40~GPa and 47~GPa and at elevated temperature ($\sim$1000~K), the $\iota$-Fe$_2$O$_3$ phase (Rh$_2$O$_3$-II type structure) can be synthetized after a temperature quench \cite{ito_determination_2009}. 
Above 54~GPa, at ambient temperature, the $\zeta$-Fe$_2$O$_3$ phase (distorted perovskite) is observed up to 67 GPa, and a metastable phase, $\theta$-Fe$_2$O$_3$ (a structure with orthorhombic symmetry), is observed up to 110 GPa. Between 57 GPa and 110 GPa, but at high temperatures ($\sim$2000~K) or after a temperature quench, the phase $\eta$-Fe$_2$O$_3$ (a post-perovskite) is stable.
In the pressure regime above 200~GPa, theoretical studies using DFT method at zero-temperature, predict solid - solid phase transitions at 233~GPa \cite{weerasinghe_computational_2015} and at 330~GPa \cite{ghosh_structural_2009}. We notice that no high-pressure melting phase transition of Fe$_2$O$_3$ has so far been reported, either experimentally nor by simulation. From a (exo-)planetology point of view, the potential presence of liquid metal oxides, which would be related to the presence of a deep conductive liquid layer in the mantle of Super-Earths which could directly contribute to the dynamo process in exoplanets \cite{nellis_possible_2012}.

In this framework, laser compression techniques are highly relevant as they enable reaching very high pressures, of the order of a few TPa \cite{duffy_ultra-high_2019}. In particular, pressures achieved using laser compression are so high that they allow constraining EOS despite large error bars. Furthermore, understanding the dynamic phase diagram of iron oxides at large strain rates typical of hypervelocity impact conditions \cite{marcus_collisional_2009}, such as meteorite impacts during accretion processes, is key to constraining composition and oxygen budget of the planet \cite{wood_accretion_2006}. In the case of Fe$_2$O$_3$, very few dynamic compression experiments have been performed, mostly due to the complexity of target design and stoichiometry control when using natural samples. Up to now, a dynamic compression dataset has been measured up to 140~GPa  along the Hugoniot using high-explosive method by McQueen et \textit{al.} \cite{mcqueen_handbook_1966} (used in \cite{liebermann_elastic_1968,anderson_shock-wave_1968} and compiled in \cite{marsh_lasl_1980}). In 1980, Kondo et al. measured the electrical resistivity of Fe$_2$O$_3$ using a double-stage light-gas gun on large natural crystals \cite{kondo_electrical_1980} and reported a transition to a low-resistivity metal-like state at 44-52~GPa pressures.

The very limited data on Fe$_2$O$_3$ at extreme pressure calls for additional Hugoniot measurements under laser shock compression (1) to extend the low-pressure data to several hundreds of GPa, (2) to obtain an EOS, possibly confirming the SESAME table, and (3) to highlight additional phase transitions by detecting anomalies in the Hugoniot curve. We present here the results of a laser shock compression experiment on Fe$_2$O$_3$ over a large range of pressures, from 20~GPa to 700~GPa. We show the EOS along the Hugoniot obtained from the velocity measurements and discuss the potential presence of phase transitions in this range along with previous studies.

\section{Method}
Experiments were conducted at the LULI 2000 high-energy laser facility. We used two laser beams at 527~nm, with either 2~ns or 5~ns square pulses and a single-pulse energy of up to 500~J. The 800~$\mu$m diameter focal spot was generated using phase plate smoothing, with a resulting maximum intensity of 1x10$^{14}$~W.cm$^{-2}$ on target when combining two laser pulses branches. Optical diagnostics such as Velocity Interferometer System for Any Reflector (VISAR) \cite{barker_laser_1972} and reflectivity measurement (which will be named Reflectivity) were used to obtain shock and particle velocities along the Hugoniot. Different types of targets were designed to perform these measurements in all the desired pressure conditions.

\begin{figure}[!h]
\center
\includegraphics[scale=0.26]{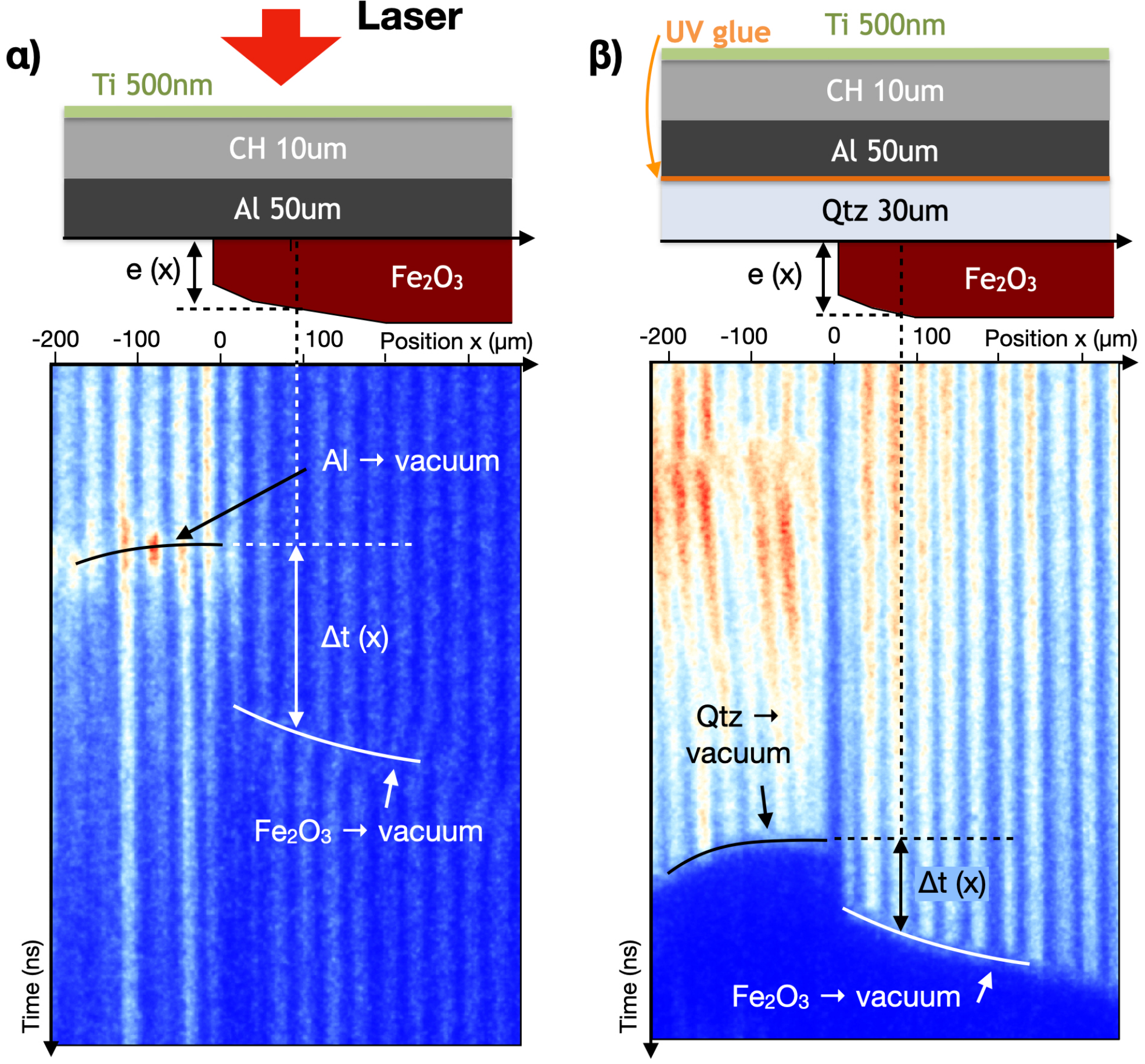}
\caption{\textbf{Target designs used to mesure $U_s$ and $U_p$ from 20~GPa to 700~GPa.}
Each target has the same ablator, which consists of a layer of parylen-N of 10 um, a layer of Al of 50 um to avoid x-rays from CH plasma heating the sample. Fe$_2$O$_3$ has a step in order to measure the shock transit time. Two different targets, $\alpha$) and $\beta$), were designed to be able to perform $U_p$ measurements in all the desired pressure conditions. An example of VISAR 2 signal is shown in the figure below each target design. On the VISAR measurements, we can clearly distinguish the breakout time profil representing the Fe$_2$O$_3$ step.
The titanium flash coating is there to avoid damage to the target due to a possible pre-pulse present in the laser pulse. 
Target design $\alpha$) was used for low pressure where the weak shock approximation is valid $U_p\approx U_{fs}/2$.
For design $\beta$), quartz was used as reference material for relative measurement of $U_p$ by impedance matching technic, for pressures above 250~GPa, for which the shock front in quartz is reflective\cite{hicks_dissociation_2006}.
}
\label{target}
\end{figure}

The target designs are shown in figure \ref{target}. They consist of a CH plastic parylen-N ablator, flashed with 500~nm Ti, and a 50 $\mu$m Al foil to avoid preheating due to x-rays produced at the ablation front. 
Target design ($\alpha$) corresponds to a Fe$_2$O$_3$ layer on half of the Al surface making a "step" target. 
Target design ($\beta$) corresponds to a Fe$_2$O$_3$ step on 30~$\mu$m quartz fixed to an Al layer with $\sim$ 2~$\mu$m thick UV glue.
Fe$_2$O$_3$ layers are polycrystalline samples obtained by Physical Vapor Deposition (PVD) and have been characterized by x-ray diffraction and energy-dispersive x-ray spectroscopy (EDS). From this analysis, we verified that our starting sample had the following characteristics: an homogeneous stoichiometry equal to the one of Fe$_2$O$_3$, and a crystallographic phase corresponding to $\alpha$-Fe$_2$O$_3$. Moreover, transmission electron microscopy (TEM) and scanning electron microscope (SEM) measurements show no particular porosity of the deposit, and the initial density was measure to be: $4.98\pm0.12$~g/cm$^3$, close to natural sample, 4.9-5.3~g/cm$^3$, and the standard value of 5.008~g/cm$^3$ used by McQueen\cite{mcqueen_handbook_1966}.
Samples also have a several hundred-nm-diameter columnar grains which are oriented along the direction of the shock propagation.
Due to target masking manufacturing, the deposited thickness was not constant near the step. To correct for this effect, we measured the exact profile thickness of the Fe$_2$O$_3$ using an interferometry profilometer (3D Optical Surface Metrology System Leica DCM8). The relative error of each measurement was around 1\% of the thickness.

On all targets, the shock velocity U$_s$ was determined from the transit time of the shock in the Fe$_2$O$_3$ step. The transit time was obtained by measuring the difference of the breakout time of the material before the Fe$_2$O$_3$ step (aluminum or quartz) and the breakout time of the Fe$_2$O$_3$ step. Breakout times were determined from reflectivity changes or velocity changes in the data.
For ($\alpha$) targets, the breakout time is measured by the change in velocity of the free surface of aluminum and Fe$_2$O$_3$, at different positions (considering 0 the position of the step; from 50~$\mu$m to 100~$\mu$m and from -50~$\mu$m to -100~$\mu$m). Note that the shock breakout is deduced from the time corresponding to the mid-height of the speed jump. On each shot, we have many data (shock breakout times and free surface velocity), from which we extracted a mean value and the standard deviation is included in the error bar.
In general, the reflectivity diagnostic was more accurate for breakout time measurements than the VISAR diagnostic, where the reflectivity was convoluted with interfringe intensity variations and timing modified due to the etalon delay. For ($\beta$) targets, the breakout time is measured by the change in reflectivity, since fringe contrast is lost at such high pressure (no free surface measurements and no windows) and no particle velocity change could be extracted.
The change in reflectivity and therefore the shock breakout time can be defined for each vertical pixel line.
By comparing the transit time in Fe$_2$O$_3$ along the sample to the step profile itself, a shock velocity can be determined for each position of the step. Finally, the shock velocity is averaged over the Fe$_2$O$_3$ region on interest, i.e. over the 50~$\mu$m to 100~$\mu$m.

Particle velocities were measured differently for each target.
Targets ($\alpha$) were designed to measure the Hugoniot in a pressure regime low enough to be in the “weak shock” limit. In this case, the free surface velocity of Fe$_2$O$_3$ is approximately the double of the fluid velocity \cite{zeldovich_shock_1967}. For iron, this approximation has been experimentally verified and yields a precision within 3 \% up to 200-300 GPa. 
This result is coherent with the definition of a weak shock, for which the energy deposited by the shock is not sufficient to melt or vaporize it. In our case, we used this target design and approximation up to 100-200~GPa, for which we are confident that we are investigating the solid phase of Fe$_2$O$_3$. The particle velocity for targets ($\beta$) was obtained using impedance matching technique with quartz as a reference (we used 7362 SESAME EOS for the quartz, equivalent to table 7360 for our pressure range). 

Shock velocity in quartz and free surface velocity of Fe$_2$O$_3$ were measured using two different line-imaging VISAR systems operating at 1.05 $\mu$m (VISAR 1) and  0.53 $\mu$m (VISAR 2). We used a time-resolved reflectivity measurement at 0.53 $\mu$m to measure changes in reflectivity, and, with the VISARs, to measure the shock transit times of the samples. The VISAR etalon velocity per fringe (VPF) associated with VISAR 1 and VISAR 2 were 12.81 and 4.96 km/s/fringe, respectively. However, the VISAR~1 etalon was 3.43 km/s/fringe for low-pressure measurements. Fringe positions in the data analysis were resolved to within 5\% of the respective VPF. Examples of VISAR~2 data are shown in Fig. 1. Depending on the 10 ns or 5 ns temporal windows, uncertainties on the breakout time were respectively 23 ps and 12 ps for the VISAR 1 and the Reflectivity streak camera, and 12 ps and 8 ps for the VISAR 2 streak camera.\\

\section{Results}

\begin{figure}[!h]
\center
\includegraphics[scale=0.65]{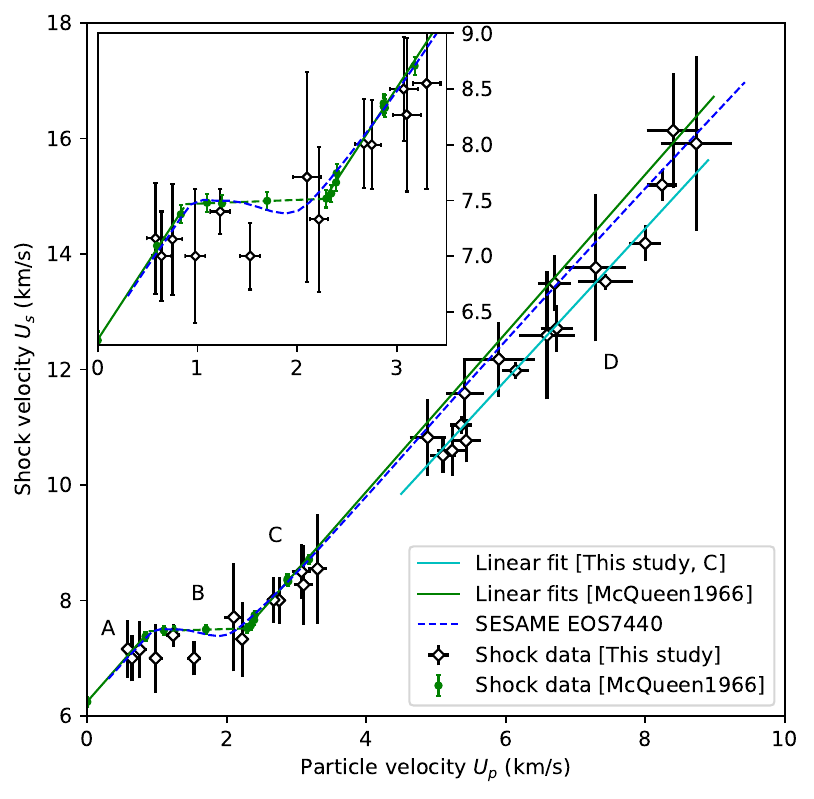}
\caption{\textbf{Hugoniot of Fe$_2$O$_3$ in the ($U_p$,$U_s$) plane.} Our shock data are represented by empty black diamond-shaped symbols. Green full circles are measurements from \cite{mcqueen_handbook_1966}. The blue dashed line represents the sesame table of hematite (7440). The cyan solid line is a linear fit ($U_s=sU_p+c_0$) of our data from $U_p=4.5$~km/s to 9~km/s. The green solid lines are a linear fit of \cite{mcqueen_handbook_1966} data. The inset focuses on the data for a particle velocity between 0 and 3.5~km/s, illustrates the transition from the low pressure phase to the high pressure phase. Experimental uncertainties are plotted in black for each points. Uncertainty in $U_s$ of the gas gun experiment data are assumed to be constant at 0.082 km/s\cite{van_thiel_compendium_1977}.}
\label{upus}
\end{figure}

Figure \ref{upus} shows the shock velocity $U_s$ versus the particle velocity $U_p$ of Fe$_2$O$_3$ determined along the principal Hugoniot. The points obtained cover the range in particle velocity from 0.6~km/s to 9~km/s. The inset corresponds to our data for particle velocities between 0 and 3.5~km/s and highlights the phase transition along the Hugoniot. Experimental uncertainties are shown in black for each point.

In our experimental data, we observe a linear behavior for particle velocity from 2~km/s to 3.6~km/s as well as from 4.5 to 9~km/s. 
Points between 0.6 and 2~km/s deviate from the general linear trend and appear to have a linear behavior with a slope coefficient near zero, forming a plateau up to $\sim$2~km/s. 
For velocities above 4.5~km/s, where no previous data exist, a linear fit of the form $U_s=sU_p+c_0$, where $s$ an c$_0$ are respectively the slope and the value at the origin, resulted in :
\begin{equation}
U_s(km/s) = 1.31(7)U_p+3.9(4)
\label{linear}
\end{equation}
The linear fit takes into account the data uncertainty by using a weighted-linear-regression model. A similar linear fit was also applied to the regions from 0.6~km/s to 2~km/s, from 2~km/s to 3.6~km/s and from 2 to 9~km/s, and the results are reported in the table \ref{linear_coef}, although they are not represented on figure \ref{upus}. Data in those two velocity regions show very good agreement with high explosive measurements obtain by \cite{mcqueen_handbook_1966}, plotted in green in figure \ref{upus}. The Hugoniot from SESAME table 7440 is plotted together with a dashed blue line.
In the 4.5 to 9~km/s region, the data are also in good agreement with the extrapolation to high pressure of \cite{mcqueen_handbook_1966} data (green line) and data from sesame table of hematite (7440).

Generally, a change of slope in the U$_s$-U$_p$ relation associated with an important volume change is often related to a phase transition. This change of the linear trend is often referred to as a plateau in the Hugoniot curve \cite{zeldovich_shock_1967}. In our low velocity data set, shown as empty diamonds symbols in figure \ref{upus}, we observe a change of the linear trend at $\sim$2~km/s. This suggests that we may have a phase transition. When considering previous datasets from Mc Queen and Marsh (green dots in figure \ref{upus}) \cite{mcqueen_handbook_1966}, the related slope change is even clearer. We can identify 3 regions of interest : a first region, labelled A, up to the first breaking point at 0.87~km/s ; then a plateau up to 2.28~km/s, labelled B, and a third region, labelled C, at higher pressure, above 2.28~km/s. Breaking point values were inferred from the crossing point of the linear fits from Mc Queen's data with the plateau. Region A and C is interpreted here as two different phases, phase A and C, and the plateau in between is interpreted as a mixed phase region. As explained by Mc Queen and Marsh \cite{mcqueen_equation_1970}, this plateau can not be strictly related to an Hugoniot measurement due to experimental possible bias from measuring the 1st shock wave arrival time only in the case of double-wave splitting from phase transitions. As a consequence, the phase C onset, measured at 2.28~km/s, can be overestimated and the phase C could be produced at lower velocities. 
At very high pressure, in the region D corresponding to data between 4.5 and 9~km/s, the linear shape and the good precision that we have on the coefficient of the slope suggests that we do not detect any phase transition with a large volume change in this region. However, we observe a general shift characterize by the c$_0$ coefficient value of equation (\ref{linear}) lower with a deviation of 10\% compared to c$_0$ of previous data \cite{mcqueen_handbook_1966} in the second phase region, who might suggest a phase transition between 3.5~km/s and 4.5~km/s.

\

\begin{figure}[!h]
\center
\includegraphics[scale=0.325]{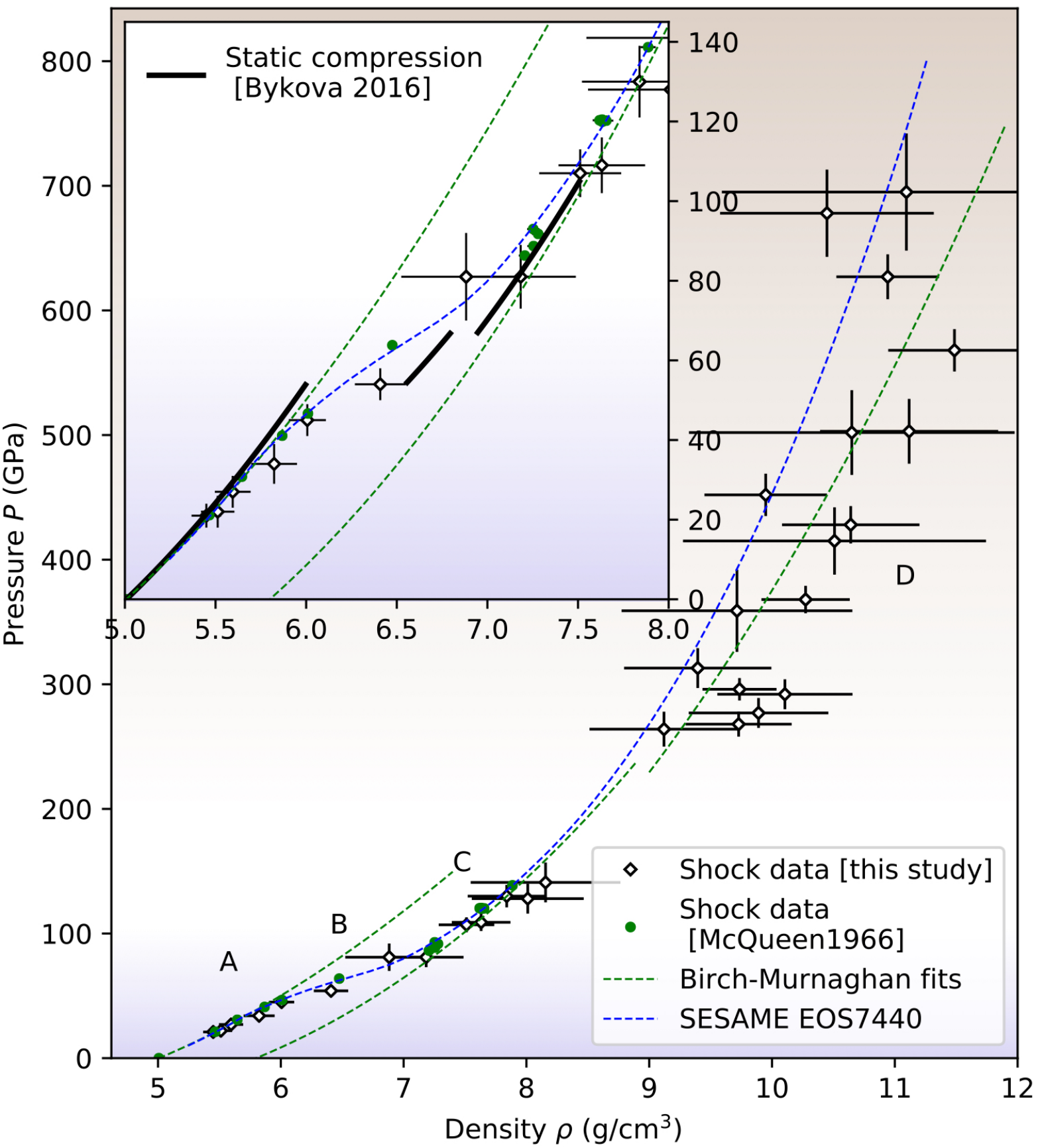}
\caption{\textbf{Hugoniot measurement of iron oxides in the ($\mathbf{\rho}$, P) plane.} Same colors as figure \ref{upus} are used for our experimental points, those of \cite{mcqueen_handbook_1966} and the SESAME EOS 7440. The black solid lines in the inset are plotted from the EOS measured using static compression \cite{bykova_structural_2016}.
Three 3rd order Birch-Murnaghan fits were performed to describe the Hugoniot in the range of 0-34~GPa; 85-140~GPa and 250-700~GPa, corresponding to region A, C, and D respectively. They are plotted as dotted green lines on the figure.}
\label{eos}
\end{figure}

Figure \ref{eos} shows pressure (P) versus density ($\rho$) of Fe$_2$O$_3$ along the principal Hugoniot. Pressures and densities are determined respectively from the Hugoniot-Rankine relation $P=\rho_0U_sU_p$, and $\rho=\rho_0U_s/(U_s-U_p)$, with $\rho_0=5.008~g/cm^3$ the standard density of Fe$_2$O$_3$ at ambient conditions. The data of Mc Queen and Marsh \cite{mcqueen_handbook_1966} are also shown on figure \ref{eos}. The isothermal equation of state, at 300~K, by Bykova et al. \cite{bykova_structural_2016} is also represented, with an ambient density of 5.008~$g/cm^3$, in order to compare directly with the shock data. The different domains of stability of phases from region A, C, and D are respectively represented by the blue, white, and brown regions.

The low-pressure phase identified above in \cite{mcqueen_handbook_1966} data corresponds to the pressure range of region A, from 0~GPa to $34.0\pm0.4$~GPa, with the pressure uncertainty obtained from the uncertainty on $U_s$ given by Van Thiel et al. \cite{van_thiel_compendium_1977}. And the high-pressure phase, corresponding to region C, is detected at $85.0 \pm0.9$~GPa. Measurements show that densities up to 11 g/cm$^3$ and pressure up to 700~GPa are reached for the highest intensity shot. 
A 3rd order Birch-Murnahan fit\cite{birch_finite_1978} was used to describe the Hugoniot in the ranges of 0-34~GPa, 85-140~GPa and 250-700~GPa, corresponding respectively to the region A, region C, and region D. The best fit parameters for phase A are $K_0 = 207~\pm~27$~GPa with $K_0' = 3.3~\pm~2.3$, and $\rho_0$ = 5.008~g/cm$^3$.
The bulk modulus is in very good agreement with published literature data for hematite namely 219~GPa \cite{bykova_structural_2016}, 206~GPa \cite{liu_static_2003}, 203~GPa \cite{liebermann_elastic_1968}, and 187~GPa and 192~GPa \cite{greenberg_pressure-induced_2018, blanchard_first-principles_2008}. All parameters are reported in table \ref{birch_coef}.

For the Birch-Murnaghan fit of region C, we used a fixed zero pressure density of 5.80~g/cm$^3$ as in \cite{anderson_shock-wave_1968}, similar to the one obtained with \textit{ab initio} calculations \cite{greenberg_pressure-induced_2018}. Results are given in table \ref{birch_coef}. The bulk modulus best fit parameter is $K_{0} = 238\pm5$~GPa, which is in good agreement with the \textit{ab initio} calculation result \cite{greenberg_pressure-induced_2018} $K_{0} = 225$~GPa for the post-perovskite high pressure phase starting at 76~GPa.
For region D, the initial density was included as a fit parameter, and the best fit parameter is $K_{85~GPa} = 471\pm21$~GPa. In comparison, fits to static experiments data \cite{bykova_structural_2016} (on the 67-100~GPa range) resulted in a bulk modulus of $K_{67~GPa} = 418\pm11$~GPa. The discrepancy at high pressure may come from non-isothermal effects due to the high temperature achieved during the shock.
It is observed that the density of the high-pressure phase is significantly higher than that predicted by the low-pressure phase equation of state. At 85~GPa, the two Birch-Murnaghan equation of state fits give a relative density (or volume) drop $(\rho_C-\rho_A)/\rho_A = 11.3\%$.

\section{Discussion}

In the following section we discuss the interpretation of phases in region A, C and D by comparing with static phase diagram of Fe$_2$O$_3$ and with a simple model for melting.

First of all, we can assume the phase in region A is the $\alpha$-Fe$_2$O$_3$ corundum structure from the ambient sample characterization (see Method section). As suggested by Mc Queen and Marsh \cite{mcqueen_handbook_1966}, this structure is expected to be stable up to 34 GPa as also reported by Anderson et al. \cite{anderson_shock-wave_1968}.
At this pressure, the plateau in the Hugoniot curve suggests that a phase transition occurs. X-ray structural measurements should be performed to confirm and identify the nature of this phase transition. Recent static compression measurements \cite{ito_determination_2009, badro_spin_2014} can help suggest possible phase transitions. While the first phase transition occurs at 54~GPa along the ambient isotherm \cite{bykova_structural_2016}, LH-DAC (Laser-Heated Diamond Anvil Cell) report the $\alpha$-Fe$_2$O$_3$ to $\iota$-Fe$_2$O$_3$ (Rh$_2$O$_3$(II)-type structure) phase transition at higher temperatures. At 34~GPa, the $\alpha$-Fe$_2$O$_3$ to $\iota$-Fe$_2$O$_3$ phase boundary temperature is at 600~K \cite{ito_determination_2009} which is in rather good agreement with the SESAME table that estimates temperatures of 435~K at the same pressure. However, the volume drop of this phase transition is only 3\% in LH-DAC studies \cite{ono_high-pressure_2004, bykova_structural_2016}. This is not in agreement with our dataset, that shows a volume drop of 11.3\% between phase in region A and region C at 85~GPa. Consequently, phase in region C would not be consistent with the $\iota$-Fe$_2$O$_3$ phase. The $\alpha$-Fe$_2$O$_3$ - $\iota$-Fe$_2$O$_3$ transition might appear at 34~GPa but should be followed by other phase transitions to explain the large volume drop observed in the shock data. 
Alternatively, the volume drop associated with the spin transition is of the same order \cite{sanson_local_2016, bykova_structural_2016}. Our shock data are in good agreement with the isotherm plot between 80 and 110~GPa (black line on the inset of figure \ref{eos}), from Bykova et al. \cite{bykova_structural_2016} who also report a spin transition.
When using the EOS from the supplementary material of Bykova et al. \cite{bykova_structural_2016}, the exact volume drop between the $\alpha$-Fe$_2$O$_3$ (here labelled phase A) and the 85~GPa high-pressure phase ($\eta$-Fe$_2$O$_3$ or  $\theta$-Fe$_2$O$_3$) is estimated to be 12.1\%.
Moreover, Kondo et al.  \cite{kondo_electrical_1980} have observed a discontinuity of the electrical resistivity along the Hugoniot between 44 and 52 GPa, suggesting metallization of hematite under shock. Kondo et al. mention the possibility of hematite undergoing a spin transition along with a Mott-insulator to metal transition. This suggests that the volume drop could be attributed to the spin transition.
Assuming this volume drop is also accompanied by a structural change seen in static compression experiments, and without considering kinetic effects, the cristalline structure of phase in region C could be either $\eta$-Fe$_2$O$_3$, $\theta$-Fe$_2$O$_3$ or a mixture of the two \cite{bykova_structural_2016}. However, as the relaxation time for those transitions is not known, only \textit{in situ} diagnostics such as diffraction could determine if one or none of those two phases is observed under shock. 

The plateau observed in region B might be interpreted as the effect of the spin transition. The spin transition at room temperature occurs over a very narrow pressure range around 50~GPa; however, the shock data indicate a softening of the spin transition ranging from 34~GPa to 85~GPa, which could be related to the temperature of the shock. Wentzcovitch et \textit{al.} \cite{wentzcovitch_anomalous_2009} have shown, using an elastic model (and in the case of the ferropericlase Mg$_{0.8125}$Fe$_{0.1875}$O), that the higher the temperature, the greater the softening. From SESAME table EOS7440, the temperature of the shock in Fe$_2$O$_3$ at our pressures is estimated to be $\sim$400~K. The pressure range of the plateau in the shock data of \cite{mcqueen_handbook_1966} and our data seem to agree with the softening effect observed in the calculation of \cite{wentzcovitch_anomalous_2009}, especially with the data of this paper corresponding to a temperature of $\sim$1000~K, were softening goes from 30~GPa to 70~GPa.

In the very high-pressure range, one could expect an additional phase transitions : a solid-liquid transition or a solid-solid transition at 233~GPa or at 330~GPa as suggested by 300~K \textit{ab initio} calculations \cite{weerasinghe_computational_2015,ghosh_structural_2009}. In the ($P,\rho$) plot figure \ref{eos}, however the large experimental uncertainty makes it difficult to observe any Hugoniot discontinuities that could be associated with such phases transitions. In our data, the typical relative uncertainty on the density is 7\%. In comparison, calculations estimate the volume drop for melting at 200-300~GPa to be 2\% for iron\cite{zhang_melting_2015}. The melting of iron occurs at a particle velocity of $\sim$3~km/s (SESAME 2140 and \cite{harmand_x-ray_2015}). If we assume the same order of magnitude for Fe$_2$O$_3$, the Hugoniot curve would cross the melting curve of Fe$_2$O$_3$ before 4.5~km/s, and in this case phase in region D could correspond to the liquid state.
Moreover, an estimation of the melting point under shock for Fe$_2$O$_3$ can be made by comparing the sesame 7440 equation of state plotted in the ($\rho$, T) plane with the melting curve given by the Lindemann model\cite{lindemann_uber_1910,poirier_introduction_2000} :
\begin{equation}
T_m = T_m^0\biggr(\frac{\rho_0}{\rho}\biggl)^{2/3}\exp\biggr(2\gamma_0\biggr(1-\frac{\rho_0}{\rho}\biggl)\biggl)
\end{equation}
with $T_m^0 = 1865 K$ and $\gamma_0 = 2.09 $\cite{liebermann_elastic_1968}, respectively the melting temperature and the Gruneïsen parameter of Fe$_2$O$_3$ at ambient pressure. The result of this comparison indicates a melting point at $\rho = 8.84~g/cm^3$, and $T = 6600~K$.
This value supports the hypothesis that phase in region D is most probably liquid Fe$_2$O$_3$. In this case, equation (\ref{linear}) would be a measure of the liquid Fe$_2$O$_3$ equation of state. 

If the shift of the Hugoniot observed at very high pressure (region D in figure \ref{upus}) with respect to extrapolation of lower pressure data is confirmed by future more precise equation of state measurements, it could be interpreted in  several ways. One possible explanation could be, that Fe$_2$O$_3$ Hugoniot crosses between 150 GPa and 250 GPa a melting curve having a negative $dT_m/dP$ slope similarly as in diamond \cite{brygoo_laser-shock_2007}. Another plausible explanation could involve a polyamorphic phase transition associated to changes in coordination number which can result in observable changes in the Hugoniot, as observed for Mg$_2$SiO$_4$ \cite{sekine_shock_2016}. Finally, dissociation reactions of Fe$_2$O$_3$, such as to Fe$_{25}$O$_{32}$ and Fe$_5$O$_7$ might be considered \cite{bykova_structural_2016, lavina_discovery_2011, lavina_unraveling_2015}. 


\begin{table}[!h]
   \center
   \caption{Comparison of $s$ and $c_0$ coefficients obtained from the linear fit on each different linear region of the $U_s-U_p$ Fe$_2$O$_3$ Hugoniot. Fits realized on McQueen et \textit{al.} data are labelled by \cite{mcqueen_handbook_1966}, and fit on laser shock data are labelled by LULI. For regions A, B, and C, we encourage using more precise coefficients resulting from fit on \cite{mcqueen_handbook_1966} data.
}
   \begin{tabularx}{\linewidth}{| c |c | X | c | c |}
     \hline
      & Data & U$_p$ range (km/s) & s & c$_0$ (km/s) \\
     \hline
     Region A & \cite{mcqueen_handbook_1966}& 0 $\rightarrow$ 0.88  & 1.37$\pm$0.05 & 6.25$\pm$0.03  \\
       \hline
     Region B & \cite{mcqueen_handbook_1966}& 0.88 $\rightarrow$ 2.28  & 0.03$\pm$0.02 & 7.44$\pm$0.03  \\
         & LULI & 0.6 $\rightarrow$ 2   & 0.1$\pm$0.4 & 7.1$\pm$0.5  \\
           \hline
     Region C& \cite{mcqueen_handbook_1966} & 2.28 $\rightarrow$ 3.2   & 1.37$\pm$0.04 & 4.4$\pm$0.1  \\
         &LULI&  2 $\rightarrow$ 3.6  & 1.1$\pm$0.5 & 5.0$\pm$1.5  \\
           \hline
     Region D & LULI & 4.5 $\rightarrow$ 9   & 1.31$\pm$0.07 & 3.9$\pm$0.4  \\
     Region C+D & LULI & 2.28 $\rightarrow$ 9  & 1.24$\pm$0.05 & 4.4$\pm$0.3  \\

     \hline
   \end{tabularx}
   \label{linear_coef} 
\end{table}

\begin{table}[!h]
   \center
   \caption{Summary of parameters obtained with a Birch-Murnaghan fit on Fe$_2$O$_3$ Hugoniot for this study, static compression study, and \textit{ab initio} calculation. These parameters correspond to the low-pressure region A, high-pressure region C and very high pressure region D. To compare the coefficients for the high-pressure phase in region C, two different fits were done, one with an initial state at ambient pressure and one at starting pressure of the high-pressure phase.}
   \begin{tabularx}{\linewidth}{|l|X|X|X|}
     \hline
     ~ & K$_0$ (GPa) & K'$_0$ & $\rho_0$(g/cm$^3$)\\
     \hline
      \textbf{Region A} & ~ & ~ & ~\\
     LULI + \cite{mcqueen_handbook_1966}   & 207$\pm$27 & 3.3$\pm$2.3 & 5.008\\
     Static \cite{bykova_structural_2016} & 219$\pm$7 & 3.5$\pm$0.4 & 5.27 \\
     Static \cite{liu_static_2003} & 206$\pm$5 & 4.3$\pm$0.3 & \ \\
     Ab. inito & 187\cite{greenberg_pressure-induced_2018}-192\cite{blanchard_first-principles_2008} & 4.1(fixed) & 4.94\\
     \hline

     \textbf{Region C} & ~ & ~ & ~\\
     LULI + \cite{mcqueen_handbook_1966}  & 238$\pm$5 & 4(fixed) & 5.80(fixed) \\
     Ab. inito & DPv:259\cite{greenberg_pressure-induced_2018} & 4.1 & 5.80 \\
     & PPv:225\cite{greenberg_pressure-induced_2018} & 4.1 & 5.84 \\
     \hline
     
    \textbf{Region D} & ~ & ~ & ~\\
     LULI & 367$\pm$18 & 4(fixed) & 6.50 \\
     \hline

   \end{tabularx}
   
    DPv : Distored-perovskite,
    PPv : Post-perovskite

   \begin{tabularx}{\linewidth}{|X|c|c|c|}
     \hline
     & K$_{HP}$ & K'$_{HP}$ & $\rho_{HP}$\\
     \hline
     LULI + \cite{mcqueen_handbook_1966}  (85~GPa) & 471$\pm$21 & 4.0 (fixed)& 7.20 \\
     Static \cite{bykova_structural_2016} (67~GPa) & 418$\pm$11 & 4.0 (fixed) & 6.95 \\
     \hline
   \end{tabularx}
   \label{birch_coef} 
\end{table}

\section*{Conclusion}
In conclusion, we have measured the equation of state of Fe$_2$O$_3$ with laser shock compression and VISAR diagnostics up to 700~GPa. We observed a phase change at low pressure that is attributed to the previously observed spin transition. Our results are in good agreement with previous gas gun measurements and consistent with the SESAME table of Fe$_2$O$_3$. Our data extend the EOS gas gun measurements on Fe$_2$O$_3$ up to pressures of 700 ~GPa and show that there is no significant volume change between 250 and 700 ~GPa. However, between 150 and 250~GPa a slight shift of the Hugoniot could suggest a phase transition with a small volume change, such as a liquid to liquid transition. 

\begin{acknowledgments}
This experiment has been realized on the LULI2000 platform, we thank all the support team and the LULI staff for their great contribution to this work. This project has received funding from the European Research Council (ERC) under the European Union’s Horizon 2020 research and innovation program (ERC PLANETDIVE grant agreement No 670787).
\end{acknowledgments}
\clearpage

\bibliographystyle{unsrt}
\bibliography{refs_luli.bib}

\end{document}